\documentclass[journal,10pt]{IEEEtran}

\usepackage{indentfirst}
\usepackage{multirow}
\usepackage{graphicx}
\usepackage{color}
\usepackage{xcolor}
\usepackage{listings}
\usepackage{enumerate}
\usepackage{subfigure}
\usepackage{amsmath,amsthm,amscd,amssymb,latexsym,upref,stmaryrd}
\usepackage[noend]{algpseudocode}
\usepackage{algorithmicx,algorithm}
\usepackage{amsfonts,hyperref,epsfig}
\usepackage{CJK}
\usepackage{float}
\usepackage{cite}
\usepackage{subfigure}
\usepackage{lipsum}
\usepackage{multicol}
\begin{document}

\title{\huge A Low-Resolution ADC Module Assisted Hybrid Beamforming Architecture for mmWave Communications}

\author{Jie Yang, Xi Yang, Shi Jin, Chao-Kai Wen, and Michalis Matthaiou}
%\vspace{-2em}
%\thanks{ IEEE TVT.}% <-this % stops a space
%\thanks{Manuscript received XXX, XX, 2017; revised XXX, XX, 2017.}

\markboth{IEEE WIRELESS COMMUNICATIONS LETTERS. VOL. XX, NO. XX, XXX 2018}%
{Submitted paper}
%\markboth{IEEE Transactions on Vehicular Technology,~Vol.~XX, No.~XX, XXX~2017}
%{Shell \MakeLowercase{\textit{et al.}}: Bare Demo of IEEEtran.cls for Journals}

\maketitle

\begin{abstract}

We propose a low-resolution analog-to-digital converter (ADC) module assisted hybrid beamforming
architecture for millimeter-wave (mmWave) communications.
%it means that the transceiver has both the low-resolution ADC and the hybrid precoding modules, a electronic switch on the transceiver switches between the low-resolution ADC and the hybrid precoding modules when training beams, while the electronic switch switches to the hybrid precoding module when transmitting data.
We prove that the proposed low-cost and flexible architecture can reduce the beam training time and complexity dramatically without degradation in the data transmission performance.
%Then we fully use the proposed system architecture to speed up the beam training process, we divide the beam training process into two phases: in phase 1, the BS switches to the hybrid precoding module and transmit beam training data, the MS switches to the low-resolution ADC module and obtain angle information; in phase 2, the MS switches to the hybrid precoding module and transmit beam training data, the BS switches to the low-resolution ADC module and obtain the rest of the angle information.
In addition, we design a fast beam training method which is suitable for the proposed system architecture. The proposed beam training method requires only $L+1$ (where $L$ is the number of paths) time slots which is smaller compared to the state-of-the-art.

\end{abstract}


%\IEEEpeerreviewmaketitle

\vspace{-1.6em}
\section{Introduction}

Communication at mmWave frequencies is viewed as a promising technique for future wireless communications. The mmWave band offers higher bandwidth than presently used sub-6GHz bands. However, the penetration loss for mmWave is large. %\cite{mmWave3}.
Hence, large antenna arrays and highly directional transmission should be combined to compensate for severe penetration loss\cite{overview}.

At the same time, the large number of antennas and high power consumption of transceivers, along with the increasing hardware complexity, induce challenges to existing beamforming techniques. Most of the traditional millimeter-wave architectures rely on analog beamforming, which is time-consuming because it can only use one beam direction at a time\cite{analogbeam}. In order to provide variable and multi-directional scanning simultaneously, hybrid beamforming system architectures have been recently investigated\cite{hy1,hy2}.
%Most importantly, hybrid beamforming system architectures can support optimal precoders to achieve larger multiplexing gains during the data transmission phase\cite{hard2,hy2}.
Note that the 802.11ad standard proposes quasi-omnidirectional beam scanning\cite{standard1}.
However, in hybrid precoding architectures, quasi-omnidirectional beam scanning creates problems to the receiver because of the high computational complexity and the requirement of phase shifters with large number of quantization bits. A low-resolution ADC architecture can reduce substantially the power consumption and computational complexity at the receiver\cite{1bit1,1bit2}.
Surprisingly, the existing literature seldom uses the low-resolution ADC assisted hybrid beamforming paradigm to speed up the beam training proceed.
%{\rl Surprisingly, the existing literature on millimeter-wave beamforming has not yet combined the low-resolution ADC paradigm with hybrid beamforming architectures.}

%Several techniques are proposed to solve those hardware constrains. (\romannumeral1) The hybrid analog/digital precoding and combining architecture is presented in works\cite{hy2}\cite{hy3}, in our work we focus on lens-based front-end. (\romannumeral2) 1-bit ADCs can reduce the resolution of ADCs at the receiver, thus lower the power consumption\cite{1bit1}.
%
%Lens has been researched for a long time.
%Lens can operate at very short pulse lengths and can scan more beamwidths than any previously known device\cite{lens1}, and lens is capable of forming low sidelobe beams\cite{lens2}.
%In \cite{beamspace} authors introduced the concept of beamspace MIMO communication and the architecture of the lens-based front-end analog beamforming for the continuous aperture phased (CAP)-MIMO transceiver.
%The work in \cite{1} first showed
%that the array response of the proposed 2D lens antenna array at the
%receiver/transmitter follows a "sinc" function, where the antenna
%with the peak response is determined by the angle of arrival
%(AoA)/departure (AoD) of the received/transmitted signal.

In this paper, we propose a low-resolution ADC module assisted hybrid beamforming architecture for mmWave communications.
%The proposed system architecture has a low-resolution ADC and a hybrid beamforming module.
An electronic switch on the proposed transceivers sweeps between the low-resolution ADC module and the hybrid beamforming module, during the beam training phase,
the transmitter switches to the hybrid beamforming module and the receiver switches to the low-resolution ADC module.
%The electronic switch on the transceivers switches to the hybrid beamforming module during the data transmission phase.
%%
For each transceiver, the hybrid beamforming module and the low-resolution ADC module will not work at the same time,
hence, the total power consumption will not increase.
In addition, we design a fast beam training method for the proposed system architecture which has two phases: in phase 1 (All-Directions-Transmitting), the mobile station (MS) calculates all the received beam directions. In phase 2 (Fine-Directions-Matching), the transmitted beams of the base station (BS) and the received beams of the MS are matched one-by-one.
The simulation results show that the proposed hardware system architecture can successfully accelerate the millimeter-wave link establishment
without degradation in the data transmission performance.
%Notation:  Upper and lower case boldface denote matrices and vectors, respectively;  the superscripts $()^{T}$ and $()^{H}$ stand for the transpose, and conjugate-transpose, respectively. The Euclidean norm and the expectation operators are denoted by $\|\cdot \|$ and $\mathbb{E}\{\cdot\}$, respectively; $z\sim \mathcal{N}(0,\sigma^2)$ denotes a real-valued Gaussian random variable z with zero mean and variance $\sigma^2$. %$\lfloor a\rfloor$ denotes the largest integer no greater than $a$.

{\bf Notations}---Throughout this paper, uppercase boldface $\mathbf{A}$ and lowercase boldface $\mathbf{a}$ denote matrices and vectors, respectively. For any matrix $\mathbf{A}$, the superscripts $\mathbf{A}^{T}$ and $\mathbf{A}^{H}$ stand for the transpose and conjugate-transpose, respectively.
For any vector $\mathbf{a}$, $\mathbf{a}^{*}$ represents the conjugate, and the 2-norm is denoted by $\|\mathbf{a}\|_{2}$.
The quantitative function is denoted by $Q(\cdot)$ and the vector operator is denoted by $\mathrm{vec}(\cdot)$. In addition, $\otimes$ represents the Kronecker product.
\vspace{-1em}

\section{System Model}
\begin{figure*}[htbp!]
\centering
\includegraphics[scale=0.78,bb= 0 0 580 180]{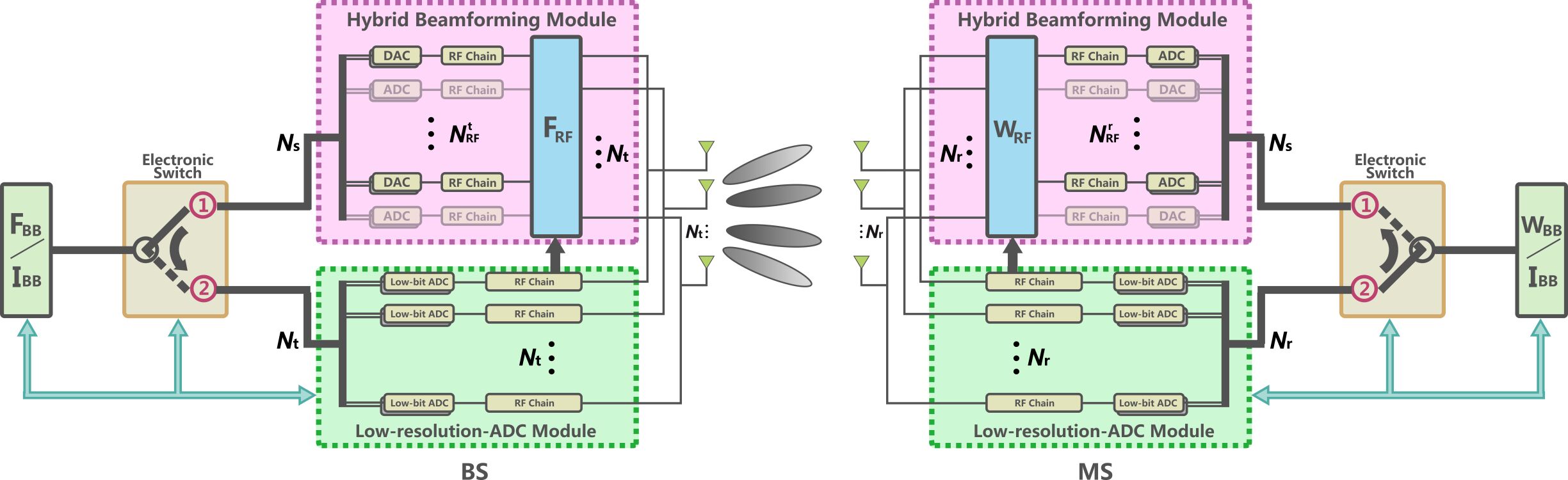}
\caption{Block diagram of the novel system architecture. The BS and MS consist of an antenna array, a low-resolution
ADC module, a hybrid beamforming module, a data interchange
module, and an electronic switch.}\label{system}
\end{figure*}
We consider a millimeter-wave system with a BS equipped with a uniform planar array (UPA) which has $N_t=N_t^{az}\times N_t^{el}$ antennas, and a MS equipped with a uniform planar array (UPA) which has $N_r=N_r^{az}\times N_r^{el}$ antennas.
We assume a narrowband point-to-point flat fading channel and obtain the uplink received data at the receive antenna array as
\begin{equation}\label{s11}
\mathbf{y}^{\tt UL}=\mathbf{H}\mathbf{x}^{\tt UL}+\mathbf{n}^{\tt UL},
\end{equation}
where $\mathbf{H}\in\mathbb{C}^{N_t \times N_r}$ denotes the uplink channel matrix between the MS and the BS.
Obviously, with the assumption of channel reciprocity, the downlink received data is given by
\begin{equation}\label{s1}
\mathbf{y}^{\tt DL}=\mathbf{H}^{T}\mathbf{x}^{\tt DL}+\mathbf{n}^{\tt DL},
\end{equation}
where $\mathbf{H}^{T}\in\mathbb{C}^{N_r \times N_t}$ represents the downlink channel matrix; $\mathbf{x}^{\tt UL}\in\mathbb{C}^{N_r \times 1}$ and $\mathbf{x}^{\tt DL}\in\mathbb{C}^{N_t \times 1}$ are the transmitted data at transmitting antenna array.
Moreover, $\mathbf{n}^{\tt UL} \sim \mathcal{CN}(0,{\sigma^2_{\tt ul}}\mathbf{I})$ and $\mathbf{n}^{\tt DL}\sim \mathcal{CN}(0,{\sigma^2_{\tt dl}}\mathbf{I})$ are Gaussian noise vectors.

Before proceeding, we first adopt the commonly used millimeter-wave channel model which characterizes the geometrical structure and limited scattering nature of millimeter-wave channels\cite{overview}. The channel is modeled as
\begin{equation}\label{s2}
\mathbf{H}^{T}=\sqrt{\frac{N_{r}N_{t}}{L}}\sum_{l=1}^{L}\alpha_{l}\mathbf{a}_{\tt Rx}(\theta_{l}^{az},\theta_{l}^{el})\mathbf{a}_{\tt Tx}^{H}(\phi_{l}^{az},\phi_{l}^{el}),
\end{equation}
where $L$ is the number of paths, $\alpha_{l}$ is the complex path gain of the $l$-th path,
$\theta_{l}^{az}$ and $\theta_{l}^{el}$ are the $l$-th azimuth and elevation angle-of-arrival (AoA) of the receiver,
$\phi_{l}^{az}$ and $\phi_{l}^{el}$ are the $l$-th azimuth and elevation angle-of-departure (AoD) of the transmitter, respectively.
$\theta_{l}^{az}$,$\theta_{l}^{el}$, $\phi_{l}^{az}$ and $\phi_{l}^{el}$ are modeled as uniformly distributed variables between $[0,2\pi)$.
$\mathbf{a}_{\tt Rx}(\theta_{l}^{az},\theta_{l}^{el})$ and $\mathbf{a}_{\tt Tx}(\phi_{l}^{az},\phi_{l}^{el})$ denote the receiving and transmitting UPA response vectors, respectively.
The expression of $\mathbf{a}_{\tt Rx}(\theta_{l}^{az},\theta_{l}^{el})$ is given by
\begin{equation}\label{s31}
\mathbf{a}_{\tt Rx}(\theta_{l}^{az},\theta_{l}^{el})=\mathbf{a}_{\tt Rx}(\theta_{l}^{az})\otimes\mathbf{a}_{\tt Rx}(\theta_{l}^{el}),
\end{equation}
where
\begin{equation}\label{s32}
\mathbf{a}_{\tt Rx}(\theta_{l}^{az}\!)\!\!=\!\!\frac{1}{\sqrt{N_r^{az}}}\!\!\left[1,e^{j\frac{2\pi d}{\lambda}\sin(\theta_{l}^{az})},\ldots,e^{j(N_r^{az}-1)\frac{2\pi d}{\lambda}\sin(\theta_{l}^{az})}\!\right]^T,
\end{equation}
and
\begin{equation}\label{s33}
\mathbf{a}_{\tt Rx}(\theta_{l}^{el})\!=\!\frac{1}{\sqrt{N_r^{el}}}\!\left[1,e^{j\frac{2\pi d}{\lambda}\sin(\theta_{l}^{el})},\ldots,e^{j(N_r^{el}-1)\frac{2\pi d}{\lambda}\sin(\theta_{l}^{el})}\!\right]^T,
\end{equation}
where $\lambda$ is the wavelength of the carrier and $d$ is the distance between two neighboring antennas.
The same procedure may be easily adapted to obtain the expression of $\mathbf{a}_{\tt Tx}(\phi_{l}^{az},\phi_{l}^{el})$.

To obtain the quantized representation of $\mathbf{H}^{T}$, we assume that the AoAs and AoDs are taken from angular grids of $G_r^{az}$, $G_r^{el}$, $G_t^{az}$ and $G_t^{el}$ points in $[0,2\pi)$, respectively.
In this paper, we assume that $G_r^{az},G_r^{el} \geq N_{RF}^r$ and $G_t^{az},G_t^{el}\geq N_{RF}^t$, and let $G_r=G_r^{az}\times G_r^{el}$ and $G_t=G_t^{az}\times G_t^{el}$, respectively. The angular grids are given by
\begin{equation}\label{s4}
\ \ \bar{\theta}_i^{az}=\arccos(2(i-1)/G_r^{az}-1),\ \ i=1,2,\ldots,G_r^{az},
\end{equation}
\begin{equation}\label{s44}
\bar{\theta}_i^{el}=\arccos(2(i-1)/G_r^{el}-1),\ \ i=1,2,\ldots,G_r^{el},
\end{equation}
\begin{equation}\label{s5}
 \ \ \bar{\phi}_j^{az}=\arccos(2(j-1)/G_t^{az}-1),\ \ j=1,2,\ldots,G_t^{az}.
\end{equation}
\begin{equation}\label{s55}
 \ \ \bar{\phi}_j^{el}=\arccos(2(j-1)/G_t^{el}-1),\ \ j=1,2,\ldots,G_t^{el}.
\end{equation}
Then, we define the receiving and transmitting dictionary matrix $\mathbf{\bar{A}}_{r}\in \mathbb{C}^{N_r \times G_r}$ and $\mathbf{\bar{A}}_{t}\in \mathbb{C}^{N_t \times G_t}$, and $\mathbf{\bar{A}}_{r}$ is given by
\begin{equation}\label{s6}
\mathbf{\bar{A}}_{r}=\mathbf{\bar{A}^{az}}_{r}\otimes \mathbf{\bar{A}^{el}}_{r},%\ \mathbf{\bar{A}}_{t}=\mathbf{\bar{A}^{az}}_{t}\otimes \mathbf{\bar{A}^{el}}_{t},
\end{equation}
where
\begin{equation}\label{s66}
\ \ \mathbf{\bar{A}^{az}}_{r}=[\mathbf{a}_{\tt Rx}(\bar{\theta}_1^{az}),\mathbf{a}_{\tt Rx}(\bar{\theta}_2^{az}),\ldots,\mathbf{a}_{\tt Rx}(\bar{\theta}_{G_r^{az}}^{az})],
\end{equation}
\begin{equation}\label{s666}
\mathbf{\bar{A}^{el}}_{r}=[\mathbf{a}_{\tt Rx}(\bar{\theta}_1^{el}),\mathbf{a}_{\tt Rx}(\bar{\theta}_2^{el}),\ldots,\mathbf{a}_{\tt Rx}(\bar{\theta}_{G_r^{el}}^{el})],
\end{equation}
%\begin{equation}\label{s67}
%\ \mathbf{A}_{t}=[\mathbf{a}_{\tt Tx}(\phi_1^{az},\phi_1^{el}),\mathbf{a}_{\tt Tx}(\phi_2^{az},\phi_2^{el}),\ldots,\mathbf{a}_{\tt Tx}(\phi_{G_t^{az}}^{az},\phi_{G_t^{el}}^{el})],
%\end{equation}
The same procedure may be easily adapted to obtain $\mathbf{\bar{A}}_{t}$.
Therefore, the quantized representation of $\mathbf{H}^{T}$ is given by
\begin{equation}\label{s8}
\mathbf{H}^{T}=\mathbf{\bar{A}}_{r}\mathbf{H}_{a}\mathbf{\bar{A}}^{H}_{t}+\mathbf{E},
\end{equation}
where $\mathbf{H}_{a}$ is the angular equivalent channel matrix, while $\mathbf{E}$ indicates the quantization error.
$\mathbf{E}$ will decrease as $G_r$ and $G_t$ increase, when $G_r$ and $G_t$ are large enough, we can assume that $\mathbf{E}=0$.
The entry index of matrix $\mathbf{H}_{a}$ represents the angular information.
For example, the $(i,j)$-th entry in $\mathbf{H}_{a}$ signifies that the azimuth and elevation AoA and AoD of this entry are $\bar{\theta}_m^{az}$, $\bar{\theta}_n^{el}$,
 $\bar{\phi}_p^{az}$ and $\bar{\phi}_q^{el}$ in (\ref{s4}), (\ref{s44}), (\ref{s5}) and (\ref{s55}), respectively, where
 $m=\lfloor \frac{i}{G_r^{el}} \rfloor$, $n=\mathrm{mod}( \frac{i}{G_r^{el}})$, $p=\lfloor \frac{j}{G_t^{el}} \rfloor$, $q=\mathrm{mod}( \frac{j}{G_t^{el}})$ . Besides the entry amplitude represents the strength of the path.
Therefore, we can find the matching beams by simply estimating $\mathbf{H}_{a}$.
%As shown in Fig. \ref{angle}, in this paper, we assume that the channel possesses two paths, therefore
%there are two large values of entries appeared in $\mathbf{H}_{a}$, and most of the other entries are 0.

%The sparsity nature of $\mathbf{H}_{a}$ motivates us to apply low-resolution ADC module in receiver side and use the compressive
%sensing technique for angles estimation.
%In the next section, we illustrate a fast beam training method used in the low-resolution ADC and hybrid beamforming combined transceiver architecture.

%\vspace{-1.2em}
\section{Proposed Transceiver Architecture}
To leverage the advantages of both the low-resolution ADC and hybrid beamforming architectures, we propose a low-resolution ADC module assisted hybrid beamforming architecture,
as illustrated in Fig. \ref{system}.
The BS and MS have the same modules: an antenna array, a low-resolution ADC module, a hybrid beamforming module, and an electronic switch.
The low-resolution ADC module and the hybrid beamforming module are connected by the electronic switch, and once one of them is switched on, the other is turned off.
Hence, the total power consumption will not increase.
Specifically, during the beam training phase,
%the BS and the MS switch continuously between the low-resolution ADC module and the hybrid beamforming module by using the electronic switch. In particular,
the transmitting terminal switches to the hybrid beamforming module and the receiving terminal switches to the low-resolution ADC module.
In the channel estimation phase and data transmission phase, the BS and MS both switch to the hybrid beamforming module.
%The parameters of the proposed system architecture are summarized in Table \uppercase\expandafter{\romannumeral1}.
We now articulate the operation of each module in detail:
%Therefore, the advantage of this architecture is that it can reduce the beam training time and complexity dramatically, besides it can promote data transmission performance.

\textbf{The antenna array} is a UPA with $N_t$ and $N_r$ antennas in the BS and MS, respectively. Each antenna array is connected to both the low-resolution ADC module and the hybrid beamforming module.

\textbf{The low-resolution ADC module} is used by the receiving terminal to obtain the AoAs and AoDs from the received beam training data during the beam training phase. The low-resolution ADC module consists of $N_t$ and $N_r$ radio frequency (RF) chains in the BS and MS, respectively. Each RF chain carries a pair of amplitude-domain in-phase and quadrature (IQ) low-resolution ADCs.
There is no digital combining matrix in the low-resolution ADC module, therefore, data from all the antennas is retained directly but with a certain degree of resolution loss.
The low-resolution ADC module and the hybrid beamforming module are not working at the same time, hence the low-resolution ADC module assisted hybrid beamforming system will not
consume much more power than the traditional hybrid beamforming system.

\textbf{The hybrid beamforming module} is used by the transmitter when transmitting beam training data during the beam training phase, or by both the transmitter and receiver during the channel estimation phase and the data transmission phase. %At the BS,
The hybrid beamforming module consists of a digital precoding (combining) matrix $\mathbf{F}_{BB}$ ($\mathbf{W}_{BB}$) and an analog precoding (combining) matrix $\mathbf{F}_{RF}$ ($\mathbf{W}_{RF}$).
The hybrid beamforming module has $N^t_{RF}$ and $N^r_{RF}$ RF chains at the BS and MS, respectively, and each RF chain carrys a pair of amplitude-domain IQ full-resolution ADCs.
The number of baseband data streams is $N_s$, with $N_s \leq N^t_{RF} \leq N_t$ and $N_s \leq N^r_{RF} \leq N_r$.
 %At the MS, the hybrid precoding module consists of  an analog combiner matrix $\mathbf{W}_{RF}\in\mathbb{C}^{N^r_{RF} \times N_r}$, a digital combiner matrix $\mathbf{W}_{BB}\in\mathbb{C}^{N_s \times N^r_{RF}}$, and $N^r_{RF}$ RF chains, each carrying a pair of amplitude-domain IQ full-resolution ADCs, where the number of baseband data streams is $N_s$ and $N_s \leq N^r_{RF} \leq N_r$.

%\textbf{The data interchange module} is used to interchange angle information between the low-resolution ADC module and the hybrid beamforming module.

\begin{figure}
\centering
\includegraphics[scale=0.32,bb= 0 0 630 150]{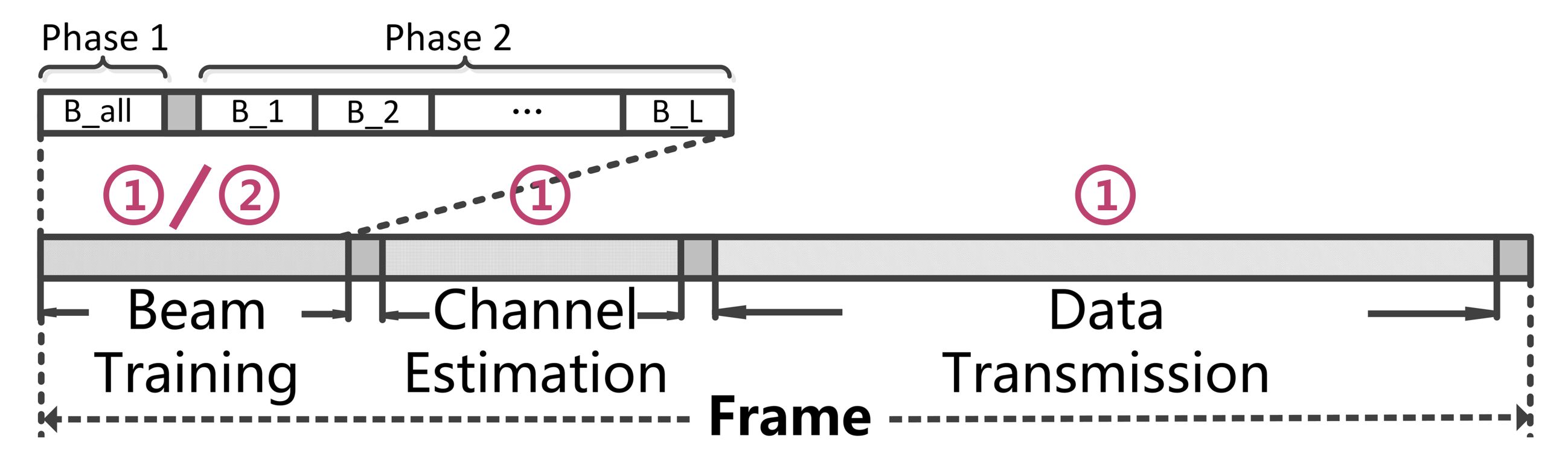}
\caption{Illustration of the frame structure: the frame consists beam training section, channel estimation section, data transmission section, and guard intervals between sections. The beam training consists two phases.}\label{frame}
\end{figure}

\textbf{The electronic switch} is strictly controlled by the frame structure as shown in Fig. \ref{frame}. The beam training process is divided into two phases: in phase 1, the BS switches to the hybrid beamforming module and transmits beam training data, the MS switches to the low-resolution ADC module and obtains angular information. In phase 2, the MS switches to the hybrid beamforming module and transmits beam training data, the BS switches to the low-resolution ADC module and obtains the rest of the angular information. During the channel estimation phase and the data transmission phase, the electronic switch switches to the hybrid beamforming module at the BS and MS.

\emph{Remark 1:} The hybrid beamforming module can provide multi-direction transmitting simultaneously,
whereas the low-resolution ADC module can capture the whole angular information rapidly within every receive operation. Therefore, these two modules coordinate with each other to considerably reduce the beam training time.
In addition, in the channel estimation and data transmission phases, the hybrid beamforming module
with full-resolution ADCs can guarantee the reliability of the data transmission and achieve large multiplexing gains.

%\begin{table}[!hbp]
%\centering
%\footnotesize
%\caption{Parameters of the proposed system architecture}
%\begin{tabular}{c|c|c|c}
%\hline
%Module & Data streams & RF chains & Antennas \\
%\hline
%BS low-resolution ADC & $N_t$ & $N_t$        & $N_t$ \\
%BS hybrid beamforming   & $N_s$ & $N_{RF}^{t}$ & $N_t$ \\
%MS low-resolution ADC & $N_r$ & $N_r$        & $N_r$ \\
%MS hybrid beamforming   & $N_s$ & $N_{RF}^{r}$ & $N_r$ \\
%\hline
%\end{tabular}
%
%\end{table}

%\vspace{-1.7em}
\section{Fast Beam Training Method}
In this section, we propose a feasible fast beam training method used in the low-resolution ADC assisted hybrid beamforming transceiver architecture.
The proposed beam training method has two phases.

\begin{figure}
\centering
\subfigure[Phase 1]{
 \label{phase:1}
 \includegraphics[scale=0.13,bb= 0 0 910 360]{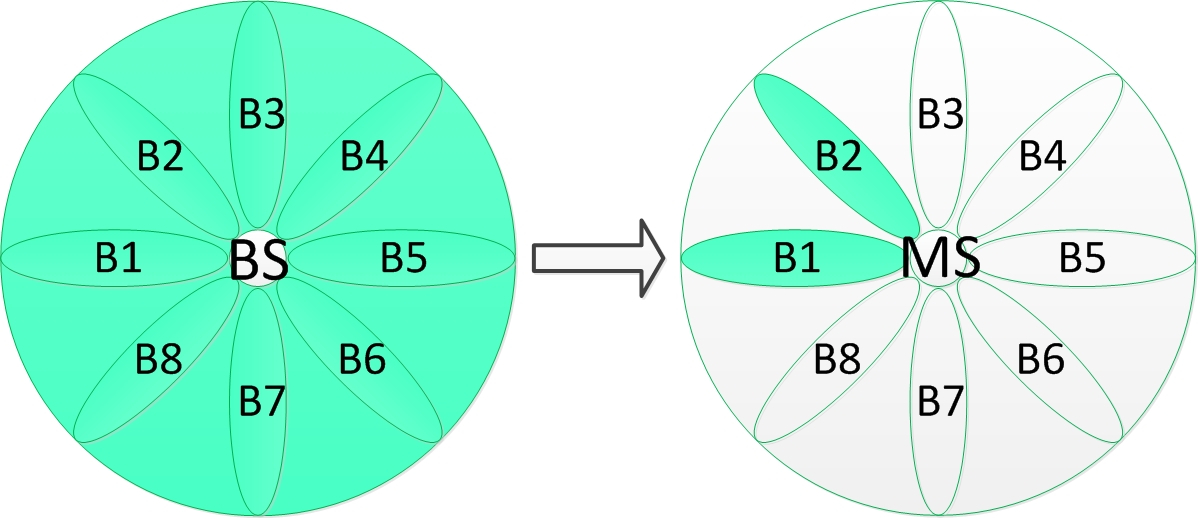}}
\hspace{0in}
\subfigure[Phase 2]{
 \label{phase:2}
 \includegraphics[scale=0.13,bb= 0 0 910 360]{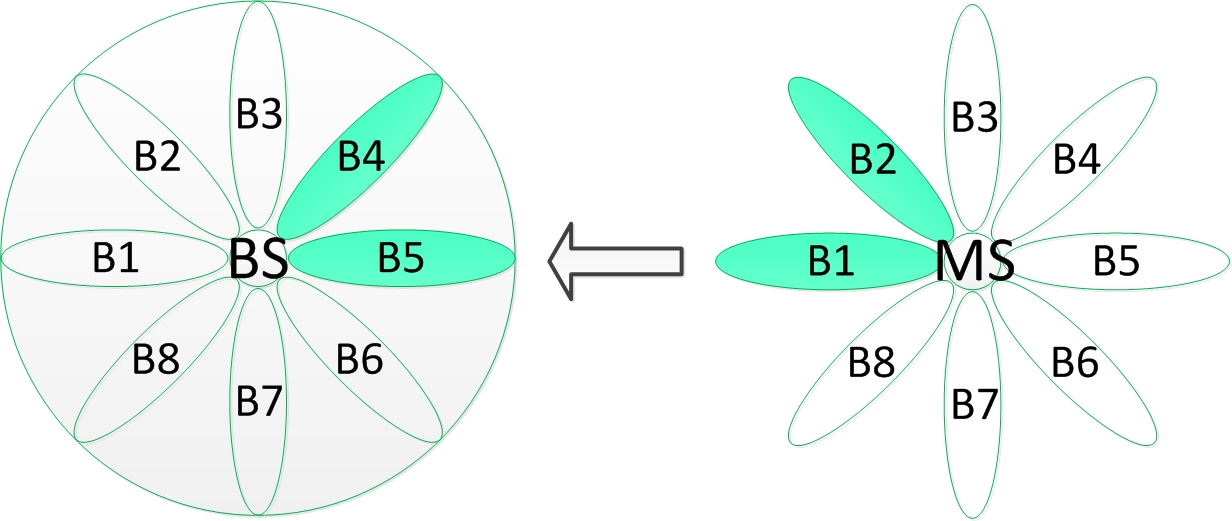}}
\caption{Illustration of two phases of beam training with $L=2$.} \label{phase}
 \end{figure}

\emph{1) Beam Training Phase 1: All-Directions-Transmitting}

During phase 1 of the beam training, the BS sends training signal across $G_t$ directions simultaneously by the hybrid beamforming module,
as shown in Fig. \ref{phase:1}.
%Since $s$ is known at the receiver side, we
%can easily remove its effect by multiplying $1/s$, Therefore, for ease of notation, we simply set $s = 1$ in this paper.
%Therefore, we simply set the training vector $\mathbf{s}=[1,1,\ldots,1]$, where $\mathbf{s}\in\mathbb{R}^{N_{RF}^t}$.
The transmitted data in (\ref{s1}) is given by
\begin{equation}\label{b1}
\mathbf{x}^{\tt DL} =  \mathbf{\bar{A}}_{t}\mathbf{1},
\end{equation}
where $\mathbf{1}\in\mathbb{R}^{G_t}$ and $\mathbf{1}=[1,1,\ldots,1]^{T}$.
Then, substituting (\ref{b1}) and (\ref{s8}) into (\ref{s1}), we obtain
\begin{equation}\label{b2}
\mathbf{y}^{\tt DL} = \mathbf{\bar{A}}_{r}\mathbf{H}_{a}\mathbf{\bar{A}}^{H}_{t} \mathbf{\bar{A}}_{t}\mathbf{1} + \mathbf{n}^{\tt DL}.
\end{equation}
Since $\mathbf{\bar{A}}^{H}_{t} \mathbf{\bar{A}}_{t}=\mathbf{I}$, we have
\begin{equation}\label{b3}
\mathbf{y}^{\tt DL} = \mathbf{\bar{A}}_{r}\mathbf{h}_v + \mathbf{n}^{\tt DL},
\end{equation}
where $\mathbf{h}_v=\mathbf{H}_{a}\mathbf{1}$.
%\begin{equation}\label{b5}
%\mathbf{y}^{\tt DL} = \mathbf{\bar{A}}_{r}\mathbf{h}_v + \mathbf{n}^{\tt DL}.
%\end{equation}
The MS switches to the low-resolution ADC module, then, the quantized receiving data can be written as
\begin{equation}\label{b4}
\mathbf{r} = Q(\mathbf{\bar{A}}_{r}\mathbf{h}_v + \mathbf{n}^{\tt DL}).
\end{equation}

It is well known from the corresponding literature that $\mathbf{h}_v$ has a sparse nature\cite{overview}.
The non-linear quantitative function $Q(\cdot)$ can be equivalent to a linear function\cite{Bussgang}.
With the knowledge of $\mathbf{\bar{A}}_{r}$ and $\mathbf{r}$, we can estimate $\mathbf{h}_v$ by the standard compressed sensing algorithm. %shown in Algorithm 1.
We denote the estimation of $\mathbf{h}_v$ as $\mathbf{h}_{v_{est}}$.
%so that $\mathbf{h}_{v_{est}}$ is a $G_r \times 1$ vector.
%According to the nature of the angular equivalent channel,
Since there are $L$ paths, there must be $L$ entries in $\mathbf{h}_{v_{est}}$ with non-vanishing power. %The largest entry is corresponding to path 1, the second largest entry is corresponding to path 2, by this analogy, the $L^{th}$ largest entry is corresponding to path L.
The indices of these $L$ entries represent the azimuth and elevation AoAs of the corresponding paths,  according to (\ref{s4}) and (\ref{s44}).
The set of AoAs is denoted as $\mathbb{S}_{AoA}=[(\theta^{az}_{AoA_{1}},\theta^{el}_{AoA_{1}}),(\theta^{az}_{AoA_{2}},\theta^{el}_{AoA_{2}}),\ldots,(\theta^{az}_{AoA_{L}},\theta^{el}_{AoA_{L}})]$.
%$\mathbb{S}_{AoA}$ is a subset of $[\theta_i:\theta_i=\arccos(2(i-1)/G_r-1),\ \ i=1,2,\ldots,G_r]$.
%The estimated AoAs are corresponding to the receiving beams shown in the right part of Fig. \ref{phase:1} with the assumption that $L=2$.

\emph{2) Beam Training Phase 2: Fine-Directions-Matching}

After obtaining the set of receiving beams $\mathbb{S}_{AoA}$, we avail of channel reciprocity, and the MS transmits the training data across the directions in set $\mathbb{S}_{AoA}$ successively in $L$ time slots by the hybrid beamforming module,
as shown in Fig. \ref{phase:2}.
The transmitted data is given by
\begin{equation}\label{b21}
\mathbf{X}^{\tt UL} =  [\mathbf{a}^{*}(\theta^{az}_{AoA_{1}},\theta^{el}_{AoA_{1}}),\ldots,\mathbf{a}^{*}(\theta^{az}_{AoA_{L}},\theta^{el}_{AoA_{L}})],
\end{equation}
where $\mathbf{X}^{\tt UL}\in\mathbb{C}^{N_r \times L}$. Then, the receiving data at the BS antenna array is given by
%\begin{equation}\label{b22}
%\mathbf{Y} = \mathbf{H}^H\mathbf{X}+\mathbf{N},
%\end{equation}
%Substituting (\ref{s8}) into (\ref{b22}), we obtain
\begin{equation}\label{b23}
\mathbf{Y}^{\tt UL} = \mathbf{\bar{A}}^{*}_{t}\mathbf{H}^{T}_{a}\mathbf{\bar{A}}^{T}_{r}\mathbf{X}^{\tt UL}+\mathbf{N}^{\tt UL},
\end{equation}
where $\mathbf{N}^{\tt UL}\sim \mathcal{N}(0,\sigma^2\mathbf{I})$ is the Gaussian noise matrix.
According to (\ref{b21}), %$\mathbf{X}$ is a sub-matrix of $\mathbf{A}_{r}$. Therefore
$\mathbf{\bar{A}}^{T}_{r}\mathbf{X}^{\tt UL}$ is a $G_r \times L$ matrix, each column of $\mathbf{\bar{A}}^{T}_{r}\mathbf{X}^{\tt UL}$ only has one non-zero entry 1 at the index of the corresponding AoA. Let us denote $\mathbf{\tilde{H}}_{a}=\mathbf{H}^{T}_{a}\mathbf{\bar{A}}^{T}_{r}\mathbf{X}^{\tt UL}$, then $\mathbf{\tilde{H}}_{a}$ is a $G_t \times L$ matrix, and each of its columns corresponds to the AoA in set $\mathbb{S}_{AoA}$.
Then, (\ref{b23}) can be simplified to
\begin{equation}\label{b233}
\mathbf{Y}^{\tt UL} = \mathbf{\bar{A}}^{*}_{t}\mathbf{\tilde{H}}_{a}+\mathbf{N}^{\tt UL}.
\end{equation}
Vectorizing $\mathbf{Y}^{\tt UL}$, we have
\begin{equation}\label{b24}
\mathbf{\tilde{y}}= \mathrm{vec}(\mathbf{Y}^{\tt UL})=(\mathbf{I}\otimes\mathbf{\bar{A}}^{*}_{t})\mathbf{\tilde{h}}_{v}+\mathbf{\tilde{n}}^{\tt UL},
\end{equation}
%\begin{equation}\label{b24}
%\begin{split}
%\mathbf{\tilde{y}} &= vec(\mathbf{Y})\\
%         &=vec( \mathbf{A}_{t}\mathbf{\tilde{H}}_{a}+\mathbf{N})\\
%         &=(\mathbf{I}\otimes\mathbf{A}_{t})\mathbf{\tilde{h}}_{a}+\mathbf{\tilde{n}},\\
%\end{split}
%\end{equation}
where $\mathbf{\tilde{h}}_{v} = \mathrm{vec}(\mathbf{\tilde{H}}_{a})$ and $\mathbf{\tilde{n}}^{\tt UL} = \mathrm{vec}(\mathbf{N}^{\tt UL})$.

The BS switches to the low-resolution ADC module, then the quantized receiving data can be written by
\begin{equation}\label{b25}
\mathbf{\tilde{r}} = Q((\mathbf{I}\otimes\mathbf{\bar{A}}^{*}_{t})\mathbf{\tilde{h}}_{v}+\mathbf{\tilde{n}}^{\tt UL}).
\end{equation}
%According to (\ref{b6}), we obtain the equivalent linear function of (\ref{b25})
%\begin{equation}\label{b26}
%\mathbf{\tilde{r}} = \mathbf{A}_e((\mathbf{I}\otimes\mathbf{A}_{t})\mathbf{\tilde{h}}_{v}+\mathbf{\tilde{n}})+ \mathbf{q}_e.
%\end{equation}
%Let $\tilde{\Phi}=\mathbf{A}_e(\mathbf{I}\otimes\mathbf{A}_{t})$ and $\mathbf{\tilde{n}}_e = \mathbf{A}_e\mathbf{\tilde{n}}+ \mathbf{q}_e$, then we obtain
%\begin{equation}\label{b26}
%\mathbf{\tilde{r}} = \tilde{\Phi}\mathbf{\tilde{h}}_{v}+\mathbf{\tilde{n}}_e.
%\end{equation}
Now, $\mathbf{\tilde{h}}_{v}$ can be estimated by the standard compressed sensing algorithm, and the estimate of $\mathbf{\tilde{h}}_{v}$ is denoted as $\mathbf{\tilde{h}}_{v_{est}}$.
Then, we reshape $\mathbf{\tilde{h}}_{v_{est}} \in\mathbb{C}^{G_t L}$ into $\mathbf{\tilde{H}}_{a_{est}}\in \mathbb{C}^{G_t \times L }$.
The index of the largest entry in the $l$-th column of $\mathbf{\tilde{H}}_{a_{est}}$ represents the AoD of path $l$. According to (\ref{s5}) and (\ref{s55}), we can find the set of AoDs denoted as  $\mathbb{S}_{AoD}=[(\phi^{az}_{AoD_{1}},\phi^{el}_{AoD_{1}}),(\phi^{az}_{AoD_{2}},\phi^{el}_{AoD_{2}}),\ldots,(\phi^{az}_{AoD_{L}},\phi^{el}_{AoD_{L}})]$.
%$\mathbb{S}_{AoD}$ is a subset of $[\phi_j:\phi_j=\arccos(2(j-1)/G_t-1),\ \ j=1,2,\ldots,G_t]$.
%The estimated AoDs are corresponding to the receiving beams shown in the left part of Fig. \ref{phase:2} with the assumption that $L=2$.
After phase 1 and phase 2 of the beam training, we finally identify the pairs of transmitting and receiving beams recorded in $\mathbb{S}_{AoD}$ and $\mathbb{S}_{AoA}$.

\vspace{-1.4em}
\section{Numerical Results}
In this section, we conduct simulations to investigate the feasibility of the proposed system architecture and the performance of the proposed beam training method.
In simulations, the number of iterations $T_{\mathrm{iter}}$ is set to 1000. The uplink SNR and downlink SNR are defined by
$\mathrm{SNR}_{\tt ul}=10\log_{10}\left(\parallel \mathbf{H}\mathbf{x}^{\tt UL} \parallel_{2}^{2}/(N_r{\sigma^2_{\tt ul}})\right)$ and $\mathrm{SNR}_{\tt dl}=10\log_{10}\left(\parallel \mathbf{H}^T\mathbf{x}^{\tt DL} \parallel_{2}^{2}/(N_t{\sigma^2_{\tt dl}})\right)$, respectively.
The function $Q(\cdot)$ is treated as the 1-bit and the 2-bit quantization functions, respectively.
$N_r^{az}$,$N_r^{el}$,$N_t^{az}$, and $N_t^{el}$ are set to 16.
 %By Bussgang decomposition \cite{Bussgang}, the non-linear quantitative function $Q(\cdot)$ can be equivalent to a linear function.
Then, the orthogonal matching pursuit algorithm in \cite{OMP} is used to estimate $\mathbf{h}_v$ in (\ref{b4}) and $\tilde{\mathbf{h}}_v$ in (\ref{b25}).
We use the successful rate
%\begin{equation*}\label{n1}
%\mathrm{SucRate} = \frac{\mathrm{Times\ of\ identifying\ the\ right\ beam\ pairs}}{T_{\mathrm{iter}}}
%\end{equation*}
to evaluate the performance of the proposed beam training method.
The successful rate is defined by the percentage of the times identifying the right beam pairs out of
the total iteration times.

\begin{figure}[htbp!]
\centering
\includegraphics[scale=0.165,bb= 0 0 1490 690]{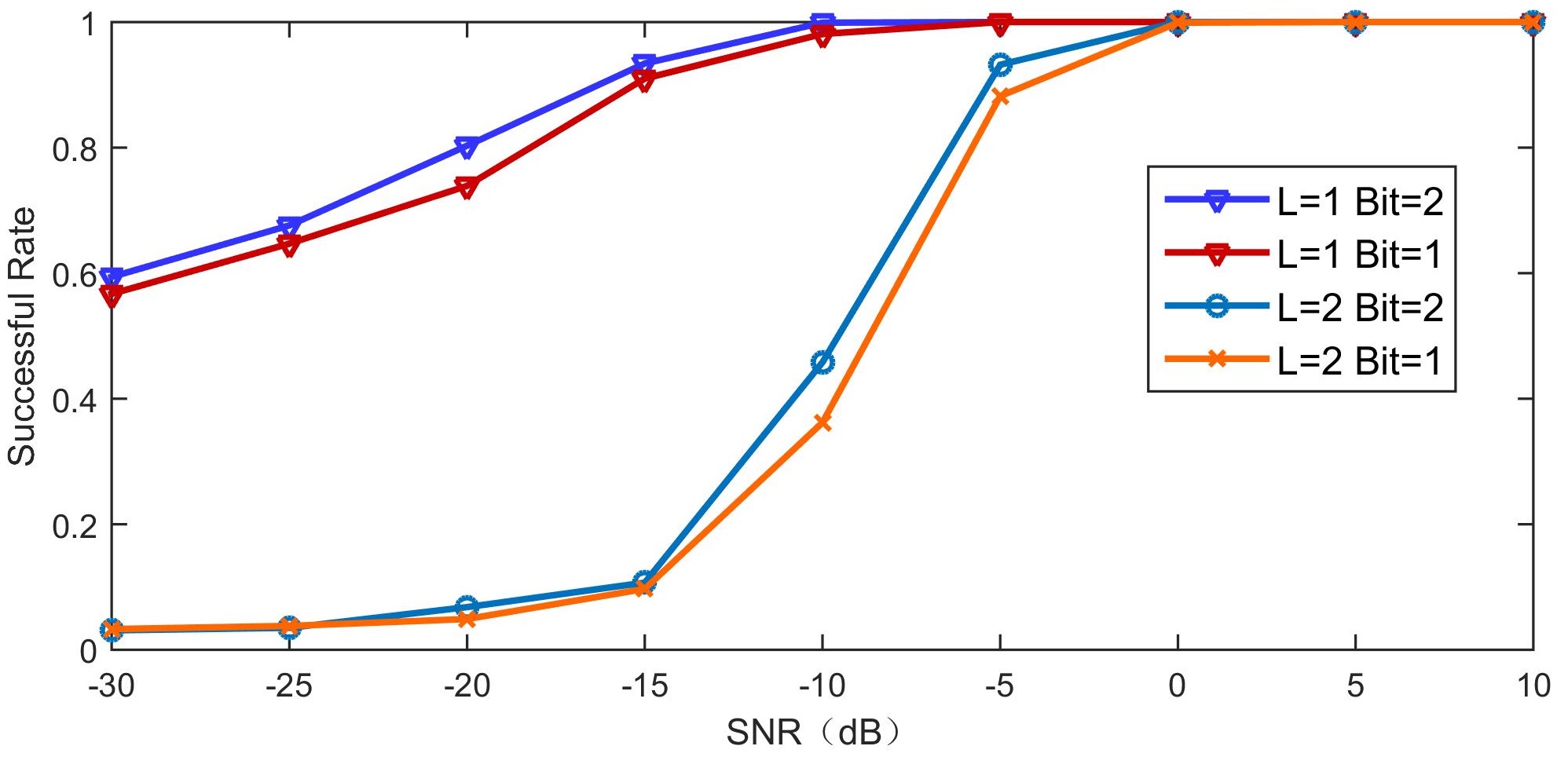}
\caption{The successful rate of the proposed beam training method with $G_r^{az}=G_r^{el}=G_t^{az}=G_t^{el}=16$.}\label{l1l2}
\end{figure}

\begin{figure}[htbp!]
\centering
\includegraphics[scale=0.165,bb= 0 0 1490 690]{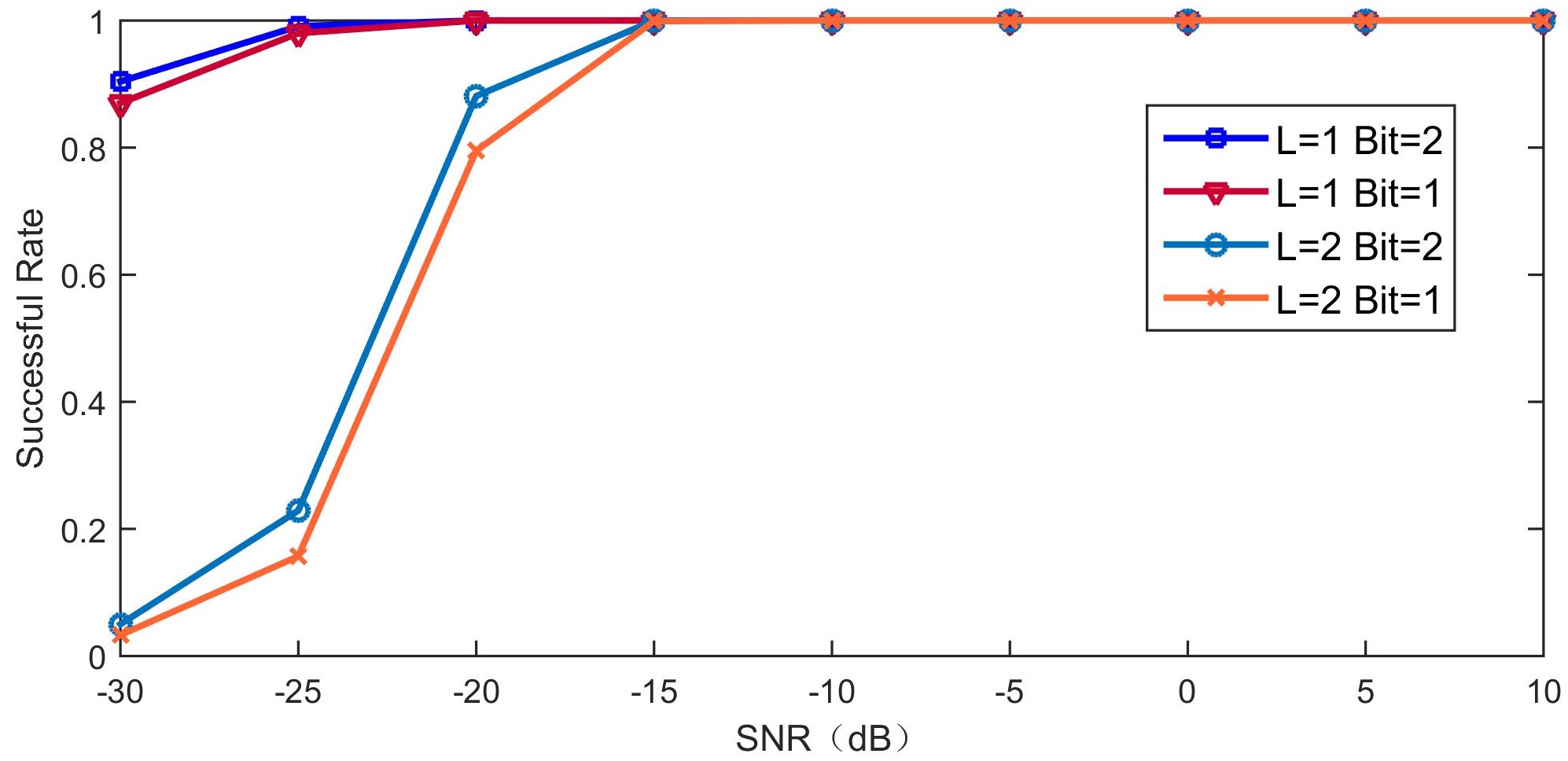}
\caption{The successful rate of the proposed beam training method with $G_r^{az}=G_r^{el}=G_t^{az}=G_t^{el}=32$.}\label{l1l2Gx2}
\end{figure}

Fig. \ref{l1l2} plots the successful rate of the proposed beam training method with 1-bit and 2-bit quantization,
and $G_r^{az}$, $G_r^{el}$, $G_t^{az}$, and $G_t^{el}$ are set to 16.
As anticipated, the performance of 2-bit quantization is better than 1-bit quantization.
In addition, the performance with $L=1$ is better than the performance with $L=2$.
When $L=2$ with 1-bit quantization, the successful rate reaches $100\%$ at SNR equal to $0 \mathrm{dB}$.
Especially, when $L=1$ with 2-bit quantization, the successful rate reaches $100\%$ at SNR equal to $-10 \mathrm{dB}$.
%The simulation results show that the proposed beam training method has extremely high successful rate when SNR is larger than $7 dB$,
Fig. \ref{l1l2Gx2} plots the successful rate of the proposed beam training method with 1-bit and 2-bit quantization,
and $G_r^{az}$, $G_r^{el}$, $G_t^{az}$, and $G_t^{el}$ are set to 32.
Compared with Fig. \ref{l1l2}, we find that with the increase of $G_r^{az}$, $G_r^{el}$, $G_t^{az}$ and $G_t^{el}$,
the successful rate of the proposed method increases.
We can infer from the simulation results that the proposed system architecture is feasible for the beam training.

\begin{table}[!hbp]
\centering
\footnotesize
\caption{beam training time comparison}
\begin{tabular}{c|c}
\hline
Beam training method               & Required time   \\
\hline
The basic method     & $(K^2\log_{K}G_t)T_{\mathrm{Slot}}$  \\
The proposed method  & $(1+L)T_{\mathrm{Slot}}$            \\
\hline
\end{tabular}
\end{table}
Before ending this subsection, we provide further discussions on the time consuming of the proposed beam training method. Let $T_{\mathrm{Slot}}$ denote the time-slot duration. The time comparison of the proposed beam training method and the basic adaptive beam training method in \cite{compare} is given in Table \uppercase\expandafter{\romannumeral2}. The basic adaptive beam training method is based on sectors scanning, where $K$ represents the number of sectors. Generally, $K=2$, $G_t=32$ and $L=2$.
From Table \uppercase\expandafter{\romannumeral2}, the proposed beam training method consumes $3 T_{\mathrm{Slot}}$, while the basic adaptive beam training method consumes $20 T_{\mathrm{Slot}}$. We can easily draw a conclusion that the proposed beam training method is more time-efficient than the existing beam training methods.

%
%\begin{figure}[htbp!]
%\centering
%\includegraphics[scale=0.12,bb= 0 0 1900 1530]{sphase2.jpg}
%\caption{The successful rate at phase 2 of the beam training.}\label{sphase2}
%\end{figure}

\vspace{-1.6em}
\section{Conclusion}
The work in this paper established a low-resolution ADC module assisted hybrid beamforming architecture for mmWave communication.
The proposed architecture was low-cost and hardware realizable by connecting low-resolution ADC modules to a hybrid beamforming architecture.
In addition, we proposed a fast beam training method employing the proposed architecture. The proposed method required only $L+1$ time slots with the assistance of a standard compressed sensing algorithm.
The simulation results verified the feasibility and the substantial reduction in time resources of the proposed architecture.
%In this paper, we design a low-resolution ADC and hybrid beamforming combined architecture for transceivers.
%The introduction of the low-resolution ADC and hybrid beamforming combined transceivers brought a new concept to the beam training, we then propose a fast beam training method under the proposed system architecture. The results show that the successful rate of the proposed method is extremely high when SNR is larger than $7 dB$. Besides, the beam training time for the proposed method is much smaller than the existing methods.
%The preceding discussions have shown that the proposed system architecture which facilitates mmWave techniques such as beam training is worth studying.
%The proposed architecture facilitates millimeter-wave techniques such as beam training.

%\section*{Acknowledgment}

\ifCLASSOPTIONcaptionsoff
  \newpage
\fi

%\begin{IEEEbiography}{Yuguang ``Michael'' Fang}
%\end{IEEEbiography}

%It is not necessary to upload the biography when you submit your manuscript.

\end{document}